\documentclass[11pt, spanish]{article}
\usepackage{cite}
\usepackage{amsmath,amsfonts,amssymb}
\usepackage[small,bf,hang]{caption}
\usepackage{slashed}
\usepackage{color}

\usepackage{hyperref}

\usepackage{ulem}

\usepackage{geometry}
\def\hybrid{
        \topmargin -20pt
        \oddsidemargin 0pt
        \headheight 0pt \headsep 0pt
        \textwidth 6.25in 
        \textheight 9.5in 
        \marginparwidth .875in
        \parskip 5pt plus 1pt \jot = 1.5ex}

\hybrid

 \csname
@addtoreset\endcsname{equation}{section}

\usepackage{epsfig}
\usepackage{color}
\usepackage{graphics}
\usepackage{graphicx}
\usepackage{slashed}

\usepackage{amsthm}
\usepackage{amssymb}
\usepackage{amsmath}
\usepackage{amsfonts}

\numberwithin{equation}{section}

\newcommand{\be}{\begin{equation}}
\newcommand{\ee}{\end{equation}}
\newcommand{\ba}{\begin{eqnarray}}
\newcommand{\ea}{\end{eqnarray}}

\newcommand{\nn}{\nonumber}

\newcommand{\un}{\underline}
\newcommand{\ov}{\overline}
\input xy
\xyoption{all}

\begin{document}
\begin{titlepage}
\rightline{}
\begin{center}
\vskip 1.5cm
 {\Large \bf{  
 $\beta$ symmetry of heterotic supergravity}}
\vskip 1.7cm

{\large\bf {Walter H. Baron$^1$,  Carmen A. N\'u\~nez$^2$} and Jes\'us A. Rodr\'iguez$^3$}
\vskip 1cm

$^1$ {\it  Instituto de F\'isica de La Plata}, (CONICET-UNLP)\\
{\it Departamento de Matem\'atica, Universidad Nacional de La Plata. }\\
{\it C. 115 s/n, B1900, La Plata, Argentina}\\
wbaron@fisica.unlp.edu.ar\\
 
\vskip .3cm

$^2$ {\it   Instituto de Astronom\'ia y F\'isica del Espacio}, (CONICET-UBA)\\
{\it Ciudad Universitaria, Pabell\'on IAFE, CABA, C1428ZAA, Argentina}\\
 carmen@iafe.uba.ar\\

\vskip .3cm

$^3$ {\it Departamento de F\'isica, Universidad de Buenos Aires.}\\
{\it Ciudad Universitaria, Pabell\'on 1, CABA, C1428ZAA, Argentina}\\jarodriguez@df.uba.ar

\vskip .1cm

\vskip .4cm

\vskip .4cm

\end{center}
\bigskip\bigskip

\begin{center} 
\textbf{Abstract}

\end{center} 
\begin{quote}  
The low energy effective action describing the  Kaluza-Klein reduction of  string theory on a $d$-torus possesses a continuous O($d, d$) global symmetry.
 The non-geometric piece of this symmetry, parameterized by a bi-vector $\beta$, was recently shown to effectively act as a hidden symmetry on the massless
 RR and universal NSNS fields of the ten dimensional parent theory, fixing their couplings.
Here we extend the analysis of this symmetry to the massless gauge and fermion fields of heterotic supergravity.  While the interactions of the  boson fields are univocally fixed by $\beta$ symmetry, we find four bilinear and two quartic   $\beta$ invariant combinations of fermions whose relative coefficients in the action must be determined by supersymmetry.  Although not fully fixed,  bilinear and quartic fermion couplings  are strongly restricted by $\beta$ symmetry at leading order in $\alpha'$.

\end{quote} 
\vfill
\setcounter{footnote}{0}
\end{titlepage}

\tableofcontents

\section{Introduction}

The low energy limit  of string theory at tree level compactified on a $d$-torus  exhibits a  continuous O($d,d; {\mathbb R}$) global symmetry  to all orders in $\alpha'$ \cite{sen}. This symmetry  is the low energy  enhancement of the discrete O($d,d; {\mathbb Z}$)  T-duality group of toroidal compactifications of string theory. Recently, it was shown that the non-geometric elements of O($d,d; {\mathbb R}$), parameterized by a  bi-vector $\beta$,  act covariantly on the  universal NSNS  \cite{bmn,bmn2} and the RR \cite{by} fields of the  ten dimensional theory, defining an effective symmetry of the uncompactified theory.  Moreover, $\beta$ transformations mix the  fields and their curvatures, in such a way that the symmetry fixes their couplings in the parent ten dimensional action. Hence, $\beta$ symmetry emerges as a useful tool  to compute higher derivative terms of the string $\alpha'$ expansion. 

In this paper we extend the analysis of $\beta$ symmetry to the gauge and fermionic sectors of heterotic supergravity. In section \ref{dft}, we obtain the $\beta$ transformation rules of the  supergravity and super Yang-Mills fields from the frame formulation of ${\cal N}=1$ supersymmetric Double Field Theory (SDFT) \cite{hk2}. We derive the action of heterotic supergravity,  up to bilinear fermion fields,   in Section \ref{action}. While the  interactions of boson fields are  univocally determined by $\beta$ symmetry,    we  find four $\beta$ invariant combinations of bilinear fermion couplings, whose relative coefficients can be fixed by supersymmetry. Interestingly, although $\beta$ symmetry does not completely determine the fermionic part of the action, it restricts the possible fermion couplings at leading order in $\alpha'$.  Quartic fermion interactions are considered in Section \ref{four}, where we find two $\beta$ invariant  combinations of quartic fermion fields, whose relative coefficients in the action can also be fixed by supersymmetry. We derive the cubic fermion terms in the supersymmetry transformations and the quartic fermion Lagrangian from SDFT. In Appendix \ref{apb} we show that the  two  quartic fermion terms of the action condense the multiple quartic fermion interactions of the ten dimensional Maxwell Einstein supergravity determined  in \cite{Bergshoeff:1981um}.  Conclusions are contained in Section \ref{conclu}. 

\section{$\beta$ transformations from heterotic double field theory}\label{dft}

We  derive the $\beta$ transformation rules of the heterotic  supergravity fields  from their action in the strong constrained and background independent frame formulation of ${\cal N}=1$ supersymmetric Double Field Theory (SDFT) \cite{hk2,jlp}. Following the formulation introduced in \cite{hk2}, we extend the procedure developed in \cite{bmn,bmn2},  to include the massless fermion and gauge fields of  heterotic supergravity. 

${\cal N}=1$ SDFT has a global $G=$O(10,10+$n_g$) symmetry and  two independent Lorentz symmetries $H_L=O(9,1)_L$ and $H_R=O(1,9+n_g)_R$, where $n_g$ is the dimension of the  gauge group. $ G, H_L$ and $H_R$ have  invariant metrics $\eta_{{M N}}$,  $\eta_{\underline {ab}}$ and $\eta_{\overline {AB}}$, respectively, with $M=(m,i), m=0,..., 19, i=1,..., n_g, \un a=0,...,9; \ov A=(\ov a,\ov i), \ov a= 0,...,9, \ov i=1,...,n_g$. 

The field content of the theory involves the following  propagating degrees of freedom: a generalized frame $E_{ M}{}^{A}$, parameterizing the coset $\frac{G}{H_L\times H_R}$, with $ A=(\un a, \ov A)$; an O$(10,10+n_g)$ scalar dilaton $d$; a  generalized gravitino $\Psi_{\ov A}$ transforming as a  Majorana-Weyl spinor of O$(9,1)_L$, as a vector of O$(1,9+n_g)_R$  and as a scalar of O$(10,10+n_g)$;  and a generalized dilatino $\rho$, transforming as a Majorana-Weyl spinor of O$(9,1)_L$ and as a scalar of O$(10,10+n_g)$. All indices are raised and lowered with their corresponding $\eta$ invariant metric. The  two Lorentz metrics  $\eta_{\underline {ab}}$ and $\eta_{\overline {AB}}$ act on the respective components $E_{M}{}^{\un a}$ and $E_{ M}{}^{\ov A}$ of the generalized frame, which  are constrained to satisfy the identities
\be
E_{{ M} \underline a} E^{ M}{}_{\underline b} = \eta_{\underline {ab}} \ , \ \ \ E_{M \underline a} E^{ M}{}_{\overline B} = 0 \ , \ \ \ E_{ M \overline A} E^{ M}{}_{\overline B} = \eta_{\overline {AB}}  \, .
\ee
The fields are formally defined on a space of $20+n_g$ dimensions and are acted on by derivatives $\partial_{ M}$, which are  strong constrained, meaning that the O(10,10+$n_g$) invariant contraction of derivatives vanishes.

Infinitesimally, O(10,10+$n_g$) acts through a constant antisymmetric matrix $h_M{}^N$, each Lorentz group through independent local antisymmetric parameters $\Lambda_{\underline {ab}}$ and $ \Lambda_{\overline {AB}}$, generalized diffeomorphisms through a generalized vector $\xi^M$ and supersymmetry through  a spinor of O(1,9)$_L$, $\epsilon$. The fields transform as
\begin{subequations}
\label{gentransf0}
\begin{align}
\delta E_{M}{}^{\un{a}} & = h_{M}{}^{N}E_{N}{}^{\un{a}} + E_{M}{}^{\un{b}}\Lambda_{\un{b}}{}^{\un{a}} +{\cal L}_\xi E_{ M}{}^{\un a}- \frac{1}{2}\ov{\epsilon}\gamma^{\un{a}}\Psi_{\ov{B}}E_{M}{}^{\ov{B}}\, , \label{gentransf0vielund}\\ 
\delta E_{M}{}^{\ov{A}} & = h_{M}{}^{N}E_{N}{}^{\ov{A}} + E_{M}{}^{\ov{B}}\Lambda_{\ov{B}}{}^{\ov{A}} +{\cal L}_\xi E_{M}{}^{\ov A}+\frac{1}{2}\ov{\epsilon}\gamma_{\un{b}}\Psi^{\ov{A}}E_{M}{}^{\un{b}}\, , \label{gentransf0vielov}\\
\delta d & ={\cal L}_\xi d - \frac{1}{4}\ov{\epsilon}\rho\, , \label{0transf}\\   
\delta\Psi_{\ov{A}} & = \Psi_{\ov{B}}\Lambda^{\ov{B}}{}_{\ov{A}} +{\cal L}_\xi\Psi_{\overline A}+ \frac{1}{4}\Lambda_{\un{bc}}\gamma^{\un{bc}}\Psi_{\ov{A}} + \nabla_{\ov{A}}\epsilon\, , \label{gentransf0gravi}\\
\delta \rho & = \frac{1}{4}\Lambda_{\un{bc}}\gamma^{\un{bc}}\rho+{\cal L}_\xi\rho - \gamma^{\un{a}}\nabla_{\un{a}}\epsilon\, , \label{gentransf0rho}
\end{align}
\end{subequations}
 up to bilinear fermion terms. ${\cal L}_\xi$ is the generalized Lie derivative, acting as
\begin{subequations}\label{gendiffeomorphisms}
\begin{align}
{\cal L}_\xi E^{ M}{}_{ A}&=\xi^{ N}\partial_{ N} E^{ M}{}_{ A}+(\partial^{ M}\xi_{ N}-\partial_{N}\xi^{ M}) E^{ N}{}_{ A}+f^{ M}{}_{ {N P}}\xi^{ N}E^{ P}{}_{ A}\,  ,\label{dife}\\
{\cal L}_\xi\Psi_{\overline A}&= \ \xi^{ M} \partial_{ M} \Psi_{\overline{A}}  \, , \\
{\cal L}_\xi d & =\ \xi^{M}\partial_{ M} d-\frac12\partial_{ M}\xi^{M}\, ,\\
   \qquad \delta_\xi \rho  & = \  {\cal L}_\xi\rho=\xi^{ M} \partial_{ M} \rho\, ,\label{diff}
\end{align}
\end{subequations}
where partial derivatives $\partial_{M}$   belong to the fundamental representation of O(10,10+$n_g$) and  the so-called fluxes or gaugings $f_{{MNP}}$  are a set of constants \cite{hk} verifying the following  linear and quadratic constraints
\be
f_{{ MNP}}=f_{[{MNP}]}\, , \qquad f_{[ {MN}}{}^{ R}f_{{P}] R}{}^{ Q}=0\, . \label{consf}
\ee
The ${\rm{O}}(9,1)_{L}$ gamma matrices can be chosen to be conventional gamma matrices in ten dimensions,  satisfying
\be
\{\gamma_{\un a},\gamma_{\un b}\}=-2\eta_{\un{ab}}\, , \label{clif}
\ee
and
the Lorentz covariant derivative is
\be
\nabla_{A}\Psi_{B} = D_{A}\Psi_{B} +w_{AB}{}^C\Psi_{C}  - \frac{1}{4}\omega_{A\un{bc}}\gamma^{\un{bc}}\Psi_{B} \, ,
\ee
with $D_{A}=\sqrt{2}E^{M}{}_{A}\partial_{M}$ and $\omega_{ABC}$  the generalized spin connection. Only the totally antisymmetric and trace parts of $\omega_{ABC}$ can be determined in terms of $E^M{}_A$ and $d$, namely
\ba
\omega_{[ABC]}&=&-E_{N[A}D_{B}E^N{}_{C]}-\frac{\sqrt2}3f_{MNP}E^M{}_AE^N{}_BE^P{}_C \, ,\label{f1}\\
 \omega_{BA}{}^B&=&-\sqrt2e^{2d}\partial_M\left(E^M{}_Ae^{-2d}\right)\, .\label{f2}
\ea

Consistency of the construction requires 
constraints which restrict the coordinate dependence of fields and gauge parameters. The strong constraint
\be
\partial_{ M} \partial^{ M} \cdots = 0 \ , \ \ \ \ \ \partial_{M} \cdots \ \partial^{ M} \cdots = 0 \, , \ \ \ \ \
f_{{MN}}{}^{ P}\partial_{ P}\cdots =0\, ,\label{StrongConstraint}
\ee
where $\cdots$ refers to products of fields, will be assumed throughout. This  locally removes   the field dependence  on $10+n_g$ coordinates, so that
fermions can be effectively defined in a $10$-dimensional tangent space.

To make contact with ${\cal N}=1$ supergravity coupled to $n_g$ vector multiplets, we split the coordinates as $x^M = (\tilde x_\mu, \,  x^\mu, x^i)$ and select  the supergravity section as a solution of the strong constraint, which simply amounts to setting $\partial_M = (0 ,\, \partial_\mu, 0)$.

The decomposition of the invariant metrics is given by
\be
\eta_{M N} = \left(\begin{matrix} 0 & \delta^\mu_\nu  &0\\ \delta^\nu_\mu & 0&0\\
0&0&\kappa_{ij}\end{matrix} \right) \ , \ \ \ \eta_{\underline {a b}} = - g_{\underline {a b}} \ , \ \ \ \eta_{\overline {ab}} = g_{\overline {ab}} \, , \ \ \ \eta_{\ov {ij}} = \kappa_{\ov {ij}}
\ee
where $\mu, \nu=0,...,9$, $g_{\underline {a b}}$ and $g_{\overline {a b}}$ are ten dimensional Minkowski metrics and $\kappa_{{ij}}=e_i{}^{\ov i}\kappa_{\ov{ij}}e_j{}^{\ov j}$ is the Killing metric of the gauge group with vielbein $e_i{}^{\ov i}$. The $\gamma^{\un a}$  matrices  verify the Clifford algebra $\{\gamma^{\un a}, \gamma^{\un b}\}=2g^{\un{ab}}$, and the non-abelian gauge sector is incorporated through the gaugings that deform the generalized Lie derivative \eqref{dife}  as
\ba
f_{MNP}=\left\{\begin{matrix}f_{ijk}& {\rm for}\ \ M, N, P = i,j,k\\
0 & {\rm otherwise}\end{matrix} \right.\, .
\ea

 We parameterize the generalized fields as 
\be
{E}_{M}{}^{\un{a}} = \frac{1}{\sqrt{2}}\left(\begin{matrix} e^{\mu\un{a}}\\
- e_{\mu}{}^{\un{a}} - C_{\rho\mu}e^{\rho\un{a}}\\
- A_{\rho i}e^{\rho\un{a}}\end{matrix}\right)  \ , \ \ \ {E}_{M}{}^{\ov{a}} = \frac{1}{\sqrt{2}}\left(\begin{matrix} e^{\mu\ov{a}}\\
e_{\mu}{}^{\ov{a}} - C_{\rho\mu}e^{\rho\ov{a}}\\
- A_{\rho i}e^{\rho\ov{a}}\end{matrix}\right) \ , \ \ \  {E}_{M}{}^{\ov{i}} = \left(\begin{matrix} 0\\
A_{\mu}{}^{i}e_{i}{}^{\ov{i}}\\
e_{i}{}^{\ov{i}}\end{matrix}\right)\, ,\label{param}
\ee
\be
e^{-2d}=\sqrt{-g}e^{-2\phi} \longrightarrow d = \phi - \frac{1}{2}\ln{\sqrt{-g}}\, , \label{gendil}
\ee
\ba
\Psi_{{A}}=(0,\Psi_{\ov{a}}, \Psi_{\ov{i}})\  &{\rm with }& \ \Psi_{\ov{a}} = \psi_{a}\delta^{a}{}_{\ov{a}}\, ,\ \ \ \  \ \Psi_{\ov{i}} = \frac{1}{\sqrt{2}}\chi_{i}e^{i}{}_{\ov{i}}\,  , \ \ \ \  \label{psi}\\
 \ \rho& =& 2\lambda + \gamma^{\mu}\psi_{\mu}\label{rho} \, , \label{parrho}
\ea
where $e_\mu{}^{\un a}, e_\mu{}^{\ov a}$ are two vielbein for the same ten dimensional metric, $g_{\mu\nu}=e_\mu{}^{\un a}g_{\un{ab}}e_\nu{}^{\un b}=e_\mu{}^{\ov a}g_{\ov{ab}}e_\nu{}^{\ov b}$, $C_{\mu\nu}=b_{\mu\nu}+\frac12A_\mu{}^iA_{\nu i}$,  $b_{\mu\nu}$ a two-form, $\phi$ is the dilaton and $\psi_a, \lambda$ and  $\chi_i$ are the supergravity  gravitino,   dilatino and  gaugino, respectively.  

Note that the derivatives $\partial_M$ belong to the fundamental representation of O(10,10+ $n_g$). In particular, the non-geometric elements of this group parameterized by a bi-vector $\beta^{\mu\nu}$ break the choice of section
\ba
\delta_\beta\partial_M\cdots=(\beta^{\mu\nu}\partial_\nu,0,0)\cdots\, ,
\ea
which must then be  imposed by the constraint
\be
\beta^{\mu\nu}\partial_\nu\cdots=0\, .\label{const}
\ee
The parameterization \eqref{param} partially fixes the double Lorentz group. Further
gauge fixing $e_\mu{}^{\ov a}\delta_{\ov a}{}^a=e_\mu{}^{\un a}\delta_{\un a}{}^a=e_\mu{}^{ a}$, one can  identify  the supergravity vielbein $e_\mu{}^a$,  i.e. $g_{\mu\nu}=e_\mu{}^ag_{ab}e_\nu{}^b$. 
Then the  components $\Lambda_{\ov{ab}}$ and $\Lambda_{\un{ab}}$ of the Lorentz parameters can be expressed in terms of the same set of indices as
\be
\Lambda_{\ov{ab}}=\delta_{\ov a}{}^a\delta_{\ov b}{}^b\ov\Lambda_{ab}\, ,\qquad  \Lambda_{\un{ab}}=\delta_{\un a}{}^a\delta_{\un b}{}^b\un\Lambda_{ab}\, .
\ee
In this way we obtain two different transformations for the gauge fixed vielbein, each one coming from the transformation of each component of the generalized frame, i.e.
\ba
\delta E^{\mu\un{a}} \quad
&\rightarrow&\quad \delta e^{\mu a}  =  - \beta^{\mu\nu}\left(e_{\nu}{}^{a}+C_{\rho\nu}e^{\rho a} \right)- e^{\mu}{}^{b}\un{\Lambda}_{b}{}^{a} - \frac{1}{2}\ov{\epsilon}\gamma^{a}\psi_{b}e^{\mu}{}^{b}\, ,\\
\delta E^{\mu}{}^{\ov{a}} \quad &
\rightarrow&\quad \delta e^{\mu}{}^{a} = \  \beta^{\mu\nu}\left(e_{\nu}{}^{a} -C_{\rho\nu}e^{\rho a} \right)+ e^{\mu}{}^{b}\ov{\Lambda}_{b}{}^{a} - \frac{1}{2}\ov{\epsilon}\gamma_{b}\psi^{a}e^{\mu b}\, ,
\ea
where we have only  considered non-vanishing constant infinitesimal antisymmetric elements $\beta^{\mu\nu}$  in  $h_{M}{}^{ N}\in $ O(10,10+$n_g$). 

These transformations must obviously coincide, which imposes the following relation between the double Lorentz parameters
\be
\ov{\Lambda}_{ab} + \un{\Lambda}_{ab} = - 2\beta_{ab} + \ov{\epsilon}\gamma_{[a}\psi_{b]}\, ,
\ee
where  $\beta_{ab}=\beta^{\mu\nu}e_{\mu a}e_{\nu b}$. Different solutions of this equation are related by redefinitions of the standard Lorentz parameter of supergravity. We choose the solution 
\be
\un{\Lambda}_{ab} = - \Lambda_{ab} - \beta_{ab} + \ov{\epsilon}\gamma_{[a}\Psi_{b]} \ , \ \ \ \ov{\Lambda}_{ab} = \Lambda_{ab} - \beta_{ab}\, ,
\ee
which leads to
\begin{subequations}\label{transfe}
\begin{align}
\delta e^{\mu}{}^{a} &= - \beta^{\mu\nu}C_{\rho\nu}e^{\rho a} + e^{\mu}{}^{b}\Lambda_{b}{}^{a} - \frac{1}{2}\ov{\epsilon}\gamma_{b}\psi^{a}e^{\mu b}\, ,\\
\delta e_{\mu}{}^{a} &= \beta^{\nu\rho}C_{\mu\rho}e_{\nu}{}^{a} + e_{\mu}{}^{b}\Lambda_{b}{}^{a} + \frac{1}{2}\ov{\epsilon}\gamma^{a}\psi_{\mu}\, .
\end{align}
\end{subequations}

These transformations and the   parameterization of the generalized dilaton \eqref{gendil}  lead to
\be
\delta\phi = \frac{1}{2}\beta^{\mu\nu}b_{\mu\nu} - \frac{1}{4}\ov{\epsilon}\rho + \frac{1}{4}\ov{\epsilon}\gamma^{\mu}\psi_{\mu}\equiv \frac{1}{2}\beta^{\mu\nu}b_{\mu\nu} - \frac{1}{2}\ov{\epsilon}\lambda\, .
\ee

To determine the remaining components of the Lorentz parameter, note that $E^{\mu\ov{i}}=0$ implies $\delta E^{\mu\ov{i}}=0$, from which we get
\be
\Lambda_{\ov{a}}{}^{\ov{i}} = \sqrt{2}\left(\beta^{\mu\nu}A_{\mu}{}^{i}e_{\nu a} + \frac{1}{4}\ov{\epsilon}\gamma_{a}\chi^{i}\right)e_{i}{}^{\ov{i}}\delta^{a}{}_{\ov{a}}\, .
\ee
The component $E_{i}{}^{\ov{i}} = e_{i}{}^{\ov{i}}=const$ requires $\delta E_{i}{}^{\ov{i}}  = \delta e_{i}{}^{\ov{i}}  = 0$, which implies
\be
\Lambda_{\ov{ij}} =\left(f_{ijk}\xi^k - \beta^{\mu\nu}A_{\mu i}A_{\nu j}\right)e^{i}{}_{\ov{i}}e^{j}{}_{\ov{j}}\, .
\ee
This partial gauge fixing of the double Lorentz group is necessary in order to recover the usual SO(1,9) Lorentz group of supergravity.

Now we can obtain the transformation rules of the remaining supergravity fields. From  $\delta E_{\mu}{}^{\ov{a}}$ in \eqref{gentransf0vielov}
we get
\be
\delta b_{\mu\nu} = - \beta^{\lambda\rho}\left(g_{\mu\lambda}g_{\rho\nu} + b_{\mu\lambda}b_{\rho\nu} - A_{\rho i}A_{[\mu}{}^{i}g_{\nu]\lambda} + \frac{1}{4}A_{\mu i}A_{\lambda}{}^{i}A_{\rho j}A_{\nu}{}^{j}\right) + \ov{\epsilon}\gamma_{[\mu}\psi_{\nu]} + \frac{1}{2}\ov{\epsilon}\gamma_{[\mu}\chi^{i}A_{\nu] i}\, ,
\ee
and from  $E_{\mu}{}^{\ov{i}}=A_{\mu}{}^{i}e_{i}{}^{\ov{i}}$ we get the following transformation rule for the gauge field:
\be\delta A_{\mu}{}^{i} = \beta^{\rho\nu}\left(g_{\mu\nu} + C_{\mu\nu}\right)A_{\rho}{}^{i} + \frac{1}{2}\ov{\epsilon}\gamma_{\mu}\chi^{i}\, .
\ee

Interestingly, as noted in \cite{cmnp}, the $\beta$ transformations of heterotic  boson fields obtained above may be recovered from the generalized diffeomorphisms \eqref{gendiffeomorphisms} with parameter $\xi^M=-\frac12\beta^{\mu\nu}\tilde x_\nu$.

For the fermion fields, parameterizing \eqref{gentransf0gravi} we get the transformation rules of the gravitino and gaugino. The former is
\be
\delta\psi_{a} = \psi_{b}\Lambda^{b}{}_{a} - \frac{1}{4}\Lambda_{bc}\gamma^{bc}\psi_{a} - \beta^{\mu\nu}\left(\psi_{b}e_{\mu}{}^{b}e_{\nu a} + \chi_{i}A_{\mu}{}^{i}e_{\nu a} + \frac{1}{4}\gamma_{\mu\nu}\psi_{a}\right) + \nabla^{(-)}_{a}\epsilon\, , 
\ee
where  
\be
\nabla^{(-)}_{a}\epsilon=\partial_{a}\epsilon + \frac{1}{4}w^{(-)}_{abc}\gamma^{bc}\epsilon
\ee
 with 
\ba
w^{(-)}_{abc}&=&e^\mu{}_a(w_{\mu bc}-\frac12H_{\mu bc})\, ,\\
H_{\mu bc}&=&3e^\nu{}_{b}e^\rho{}_{c}\left(\partial_{[\mu}b_{\nu\rho]}-A_{[\rho}{}^i\partial_{\mu}A_{\nu]i}+\frac13f_{ijk}A_\mu{}^iA_\nu{}^jA_\rho{}^k\right)\, .
\ea 

The  gaugino transforms as
\be
\delta\chi^{i} = \beta^{\mu\nu}\left(2A_{\mu}{}^{i}\psi_{\nu} + A_{\mu}{}^{i}A_{\nu}{}^{j}\chi_{j} - \frac{1}{4}\gamma_{\mu\nu}\chi^{i}\right) - \frac{1}{4}\Lambda_{ab}\gamma^{ab}\chi^{i} -f_{ijk}\chi^j\xi^k- \frac{1}{4}F_{\mu\nu}{}^{i}\gamma^{\mu\nu}\epsilon\, ,
\ee
where
\be
F_{\mu\nu}{}^i=2\partial_{[\mu}A_{\nu]}{}^i-f^i{}_{jk}A_\mu{}^jA_\nu{}^k\, .
\ee

Finally, the parameterization of \eqref{gentransf0rho} leads to the following transformation rule of the generalized dilatino:
\be
\delta\rho = - \frac{1}{4}\Lambda_{ab}\gamma^{ab}\rho - \frac{1}{4}\beta^{\mu\nu}\gamma_{\mu\nu}\rho + \gamma^{\mu}\nabla_{\mu}\epsilon - \frac{1}{24}H_{abc}\gamma^{abc}\epsilon - \partial_{\mu}\phi\gamma^{\mu}\epsilon \, .
\ee

Summarizing, we have derived the leading order transformation rules of the heterotic supergravity fields  from a gauge fixing of the duality covariant components of the generalized fields in ${\cal N}=1$ supersymmetric DFT. In particular, the $\beta$  transformations are
\begin{subequations}\label{transfe}
\begin{align}
\delta_\beta e_{\mu}{}^{a} &= \beta^{\nu\rho}C_{\mu\rho}e_{\nu}{}^{a} \, ,\qquad \delta_\beta\phi = \frac{1}{2}\beta^{\mu\nu}b_{\mu\nu}\, , \label{transfedil}\\
\delta_\beta b_{\mu\nu} &= - \beta^{\lambda\rho}\left(g_{\mu\lambda}g_{\rho\nu} + b_{\mu\lambda}b_{\rho\nu} - A_{\rho i}A_{[\mu}{}^{i}g_{\nu]\lambda} + \frac{1}{4}A_{\mu i}A_{\lambda}{}^{i}A_{\rho j}A_{\nu}{}^{j}\right) \, ,\\
\delta_\beta A_{\mu}{}^{i} &= \beta^{\rho\nu}\left(g_{\mu\nu} + C_{\mu\nu}\right)A_{\rho}{}^{i} \, ,\\
\delta_\beta\psi_{a}& =  - \beta^{\mu\nu}\left(\psi_{b}e_{\mu}{}^{b}e_{\nu a} + \chi_{i}A_{\mu}{}^{i}e_{\nu a} + \frac{1}{4}\gamma_{\mu\nu}\psi_{a}\right) \, , \label{transfepsi}\\
\delta_\beta\rho &= - \frac{1}{4}\beta^{\mu\nu}\gamma_{\mu\nu}\rho  \, ,\label{transferho}\\
\delta_\beta\chi^{i}& = \beta^{\mu\nu}\left(2A_{\mu}{}^{i}\psi_{\nu} + A_{\mu}{}^{i}A_{\nu}{}^{j}\chi_{j} - \frac{1}{4}\gamma_{\mu\nu}\chi^{i}\right)\, ,\label{transfechi}
\end{align}
\end{subequations}
and are subjected to the constraint $\beta^{\mu\nu}\partial_\nu\cdots=0$.

\section{$\beta$ symmetry of heterotic supergravity}\label{action}

In this section we derive the invariant action  under the $\beta$ transformation rules obtained above, up to bilinear fermion terms. 
We show that requiring $\beta$ invariance of the most general combination of diffeomorphism, gauge and Lorentz invariant terms univocally fixes the couplings of the boson fields to those of  heterotic supergravity. Instead, in the fermionic sector, we find four  $\beta$ invariant  combinations of bilinear fermion terms, and we fix their relative coefficients   by supersymmetry.

\subsection{Boson fields}

Consider the most general combination of  gauge  invariant  terms involving only boson fields, i.e.\footnote{Note that while we denote a single scalar function $f(x)$ here, a more general combination consistent with gauge symmetries can include different functions  $f_i(x)$  in front of each of the five couplings within the brackets. This generalization is straightforward, as each function will satisfy the same equation as $f(x)$ in equation (\ref{varSB}).}
\be
S_{\rm B}=\int d^{10}x\sqrt {-g}f(\phi)L=\int d^{10}x\sqrt {-g}f(\phi)\left[R+a\square\phi+b(\nabla\phi)^2+cH_{abc}H^{abc}+dF^i_{ab}F_i^{ab}\right]\, , \label{genericaction}
\ee
where
$
H_{abc}=e^\mu{}_aH_{\mu bc}$ and $F^i_{ab}=e^\mu{}_ae^\nu{}_bF^i_{\mu\nu}.
$

Using   \eqref{transfe}  and the constraint \eqref{const}, we get the following $\beta$ transformations 
\begin{subequations}\label{transfcurv}
\begin{align}
\delta_\beta\sqrt{-g}&=\sqrt{-g}\beta^{ab} b_{ab}\, ,\\
\delta_\beta w_{abc} &= \beta_{[b}{}^{d}\left(H_{c]ad} + F_{c]a}{}^{i}A_{d i}\right) - \frac{1}{2}\beta_{a}{}^{d}\left(H_{bcd} + F_{bc}{}^{i}A_{d i}\right)\, ,\\
\delta_\beta R&=-2\nabla_a\left[\left(H^a{}_{bc}+A_{bi}F_{c}{}^{ai}\right)\beta^{bc}\right]-\frac12\nabla_a\beta^{bc}\left(H^a{}_{bc}+2F^a{}_{bi}A_{c}{}^i-F_{bc}{}^iA^a{}_{i}\right)\, ,\\
\delta_\beta(\nabla\phi )^2&= \beta^{bc}\left(H^a{}_{bc} + F^a{}_{b}{}^{i}A_{ci}\right)\nabla_a\phi\, , \label{fi2}\\
\delta_\beta\square\phi&=\frac12\nabla_a\left[\left(H^a{}_{bc}+F^a{}_{bi}A_c{}^i\right)\beta^{bc}\right]+\nabla_a\phi\beta^{bc}\left(H^a{}_{bc}+F^a{}_{bi}A_c{}^i\right)\, ,\label{sfi}\\
\delta_\beta F_{ab}{}^{i} &= 2\beta_{[a}{}^{c}F_{b]c}{}^{i} + \left(\nabla^{c}\beta_{ab} - 2\nabla_{[a}\beta_{b]}{}^{c} + \beta^{cd}\left(H_{abd} + F_{ab}{}^{j}A_{dj}\right)\right)A_{c}{}^{i}\, ,\\
\delta_\beta H_{abc} & =  - 3\nabla_{[a}\beta_{bc]} - 3\beta^{d}{}_{[a}F_{bc]}{}^{i}A_{di}\, ,
\end{align}
\end{subequations}
which determine  the following variation of the action \eqref{genericaction}
\ba
\delta_\beta S_{\rm B}&=&\int d^{10}x\sqrt {-g}f(\phi)\left\{-2\left(1-\frac14b\right)\nabla_a\left[\left(H^a{}_{bc}+A_{bi}F_c{}^{ai}\right)\beta^{bc}\right]\right.\nn\\
&&\ \ \ \ \ \ \ \ \ \ \ \ \ \ \ \ \ \ \ \ \ \ \ \ +(a+b)\nabla_a\phi\beta^{bc}\left(H^a{}_{bc}+F^a{}_{bi}A_c{}^i\right)-\frac12(1+12c)\nabla_a\beta_{bc}H^{abc}\nn\\
&&\ \ \ \ \ \ \ \ \ \ \ \ \ \ \ \ \ \ \ \ \  \ \ \ -\frac12(1+4d)\nabla_a\beta_{bc}\left(2F^{abi}A^c{}_i-A^{ai}F^{bc}{}_i\right)\nn\\
&& \ \ \ \ \ \ \ \ \ \ \ \ \ \ \ \ \ \ \ \ \ \ \ \  \left.+2(d-3c)\beta^d{}_aF_{bc}{}^iA_{di}H^{abc}\right\}\nn\\
&& +\frac12\int d^{10}x\sqrt{-g}\beta^{ab}b_{ab}\left(2f(\phi)+f'(\phi)\right)\left[R+a\square\phi+b(\nabla\phi)^2+cH^2+dF^2\right]\, \label{varSB}
\ea

Therefore, $\beta$ invariance  requires
\ba
a=-b=-4\, ,\qquad c=-\frac1{12}\, ,\qquad d=-\frac14\, ,\qquad f(\phi)=e^{-2\phi}\, ,
\ea
and leads to
\ba
S_{\rm B}=\int d^{10}x\sqrt {-g}e^{-2\phi}\left[R+4\square\phi-4(\nabla\phi)^2-\frac1{12}H_{\mu\nu\rho}H^{\mu\nu\rho}-\frac14F_{\mu\nu}F^{\mu\nu}\right]\, , \label{sb}
\ea
 precisely the leading order action of the bosonic sector of  heterotic supergravity. This result completes  the analysis in \cite{bmn,bmn2} with the addition of the gauge sector.

\subsection{Fermion fields}

To obtain the $\beta$ invariant fermionic piece of the action, we consider all possible gauge invariant  bilinear fermion couplings, i.e.
\ba
F_{1} &=& \ov{\rho}\gamma^{a}\nabla_{a}\rho\, , \qquad \ \ \ \ F_{2} = \ov{\psi}_a\gamma^{abc}\nabla_{b}\psi_c\, , \qquad F_{3} = 
\ov{\psi}_a\gamma^{a}\nabla^{b}\psi_b\, , \qquad F_{4} = \ov{\psi}_a\gamma^{b}\nabla^{a}\psi_b\, , \nn\\
  F_{5} &=& \ov{\psi}^a\gamma^{b}\nabla_{b}\psi_a\, , \ \qquad F_{6} = \ov{\chi}_i\gamma^{a}\nabla_{a}\chi^i\, , \ \ \qquad F_{7} = \ov{\rho}\gamma^{ab}\nabla_{a}\psi_{b}\, , \qquad \ F_{8} = \ov{\rho}\nabla^{a}\psi_{a}\, , \nn\\
 F_{9}& =& \ov{\psi}_{a}\gamma^{ab}\nabla_{b}\rho\, , \ \qquad F_{10} = \ov{\psi}_{a}\nabla^{a}\rho\, ,\nn\\
B_{1}^{\phi} &=& \ov{\psi}_b\gamma^{b}\psi^{a}\nabla_{a}\phi\, , \ \
\quad B_{2}^{\phi} = \ov{\rho}\gamma^{ab}\psi_{a}\nabla_{b}\phi\, , 
\qquad  B_{3}^{\phi} = \ov{\rho}\psi^{a}\nabla_{a}\phi\, ,\nn\\
B_{1}^{H} &=& \ov{\rho}\gamma^{abc}\rho H_{abc}\, , \quad \ \  B_{2}^{H} = \ov{\chi}^i\gamma^{abc}\chi_i H_{abc}\, , \quad B_{3}^{H} = \ov{\psi}_a\gamma^{abcde}\psi_b H_{cde}\, , \quad B_{4}^{H} = \ov{\psi}_d\gamma^{abc}\psi^d H_{abc}\, , \nn\\
B_{5}^{H} &=& \ov{\psi}^a\gamma^{bcd}\psi_d H_{abc}\, , \ \  B_{6}^{H} = \ov{\psi}^a\gamma^{b}\psi^{c}H_{abc}\, , \quad \ B_{7}^{H} = \ov{\rho}\gamma^{abcd}\psi_{d}H_{abc}\, ,\quad \ \ \
B_{8}^{H} = \ov{\rho}\gamma^{ab}\psi^c H_{abc}\, ,\nn\\
B_{1}^{F} &=& \ov{\rho}\gamma^{ab}\chi_{i}F_{ab}{}^{i}\, ,\qquad B_{2}^{F} = \ov{\psi}_c\gamma^{abc}\chi_{i}F_{ab}{}^i\, ,\quad B_{3}^{F} = \ov{\psi}^a\gamma^{b}\chi_{i}F_{ab}{}^{i}\, ,\label{bilinear}
\ea
where 
\ba
\nabla_a\rho&=&\partial_a\rho+\frac14w_{abc}\gamma^{bc}\rho\, ,\\
\nabla_a\psi_b&=&\partial_a\psi_b+w_{ab}{}^c\psi_c+\frac14w_{abc}\gamma^{bc}\rho\, ,\\
\nabla_a\chi^i&=&\partial_a\chi^i+\frac14w_{abc}\gamma^{bc}\chi^i-A_a^j\chi^kf^i{}_{jk}\, .
\ea

Actually, we should include a term $\sqrt{-g} f_i(\phi)$ in front of each coupling in the list (\ref{bilinear}). However, the same reasoning from the previous subsection applies, ultimately resulting in an overall invariant factor of $\sqrt{-g} e^{-2\phi}$, which we will omit for brevity.

The following $\beta$ transformations will be useful:
\ba
\delta_\beta\left(\nabla_a\rho\right)&=&-\frac14\nabla_a\left(\beta^{bc}\gamma_{bc}\rho\right)-\frac18\left[\left(H_{dbc}+A_{di}F_{bc}^i\right)\beta_a{}^d-2\left(H_{dab}+A_{di}F_{ab}^i\right)\beta_c{}^d\right]\gamma^{bc}\rho\, ,\ \ \ \ \ \  \label{nrho}\\
\delta_\beta\left(\nabla_a\psi_b\right)&=&\nabla_a\left[\beta_{bc}\left(\psi^c+A^c{}_i\chi^i\right)-\frac14\beta^{cd}\gamma_{cd}\psi_b\right]\nn\\
&&-\frac12\left[\left(H_{dbc}+A_{di}F_{bc}^i\right)\beta_a{}^d-2\left(H_{da[b}+A_{di}F_{a[b}^i\right)\beta_{c]}{}^d\right]\psi^{c}\, \nn  \\
&&-\frac18\left[\left(H_{ecd}+A_{ei}F_{cd}^i\right)\beta_a{}^e-2\left(H_{eac}+A_{ei}F_{ac}^i\right)\beta_{d}{}^e\right]\gamma^{cd}\psi_{b}\, ,\\
\delta_\beta\left(\nabla_a\chi^i\right)&=&\nabla_a\left[\beta^{bc}\left(2A_b{}^i\psi_c+A_b{}^iA_c{}^j\chi_j\chi^i-\frac14\gamma_{bc}\chi^i\right)\right]\nn\\
&&-\frac18\left[\left(H_{dbc}+A_{dj}F_{bc}^j\right)\beta_a{}^d-2\left(H_{da[b}+A_{di}F_{a[b}^i\right)\beta_{c]}{}^d\right]\gamma^{bc}\chi^i\, .
\ea

Interestingly, not all the bilinear fermion fields \eqref{bilinear} are allowed in a $\beta$ symmetric action. 
Indeed, a general inspection reveals the following structure
\be
\delta_\beta T=\delta_{QC}T+\delta_{NC}T\, ,
\ee
where $T$ denotes any  bilinear fermion coupling built in terms of $\rho, \psi^a$ and $\chi^i$. $\delta_{QC}T$ is a quasi-covariant transformation
and $\delta_{NC}T$ is a non-covariant transformation containing couplings that differ from those in $T$.
In particular, using \eqref{transfepsi}, the structure $T^{ab}{}_{c_1...c_n}=\ov\psi^a\gamma_{c_1...c_n}\psi^b$ transforms as the sum of the following two terms
\ba
\delta_{QC} T^{ab}{}_{c_1...c_n}&=&\beta^a{}_dT^{db}{}_{c_1...c_n}+\beta^b{}_dT^{ad}{}_{c_1...c_n}-\beta_{c_1}{}^dT^{ab}{}_{dc_2...c_n}-\cdots-\beta_{c_n}{}^dT^{ab}{}_{c_1...d}\;, \;\;\;\;\label{bitransf}\\
\delta_{NC} T^{ab}{}_{c_1...c_n}&=&\beta^{ad}A_d{}^i\ov\chi_i\gamma_{c_1...c_n}\psi^b+\beta^{bd}A_d{}^i\ov\psi^a\gamma_{c_1...c_n}\chi_i\, ,
\ea
where we have used \eqref{i1}.
Note that if the vector index of one of the gravitini ($a$ or $b$) is contracted with  the vector index of a $\gamma$ matrix, the corresponding $\beta$ transformations in \eqref{bitransf} add up and cannot be canceled by variations of any other terms. Indeed, terms in which the vector index of a gravitino or of a $\gamma$ matrix is contracted with that of a derivative or  a boson field,  transform to expressions in which one index of $\beta$ is contracted with the gravitino or the $\gamma$ matrix and the other one with the boson field.  Therefore, $\beta$ symmetry does not allow the contraction of the vector indices of the gravitini with those of the gamma matrices,   and the couplings $F_2, F_3, F_4, F_7, F_9, B_1^\phi, B_2^\phi, B_3^H, B_5^H$ and $B_7^H$ cannot belong to a $\beta$ invariant Lagrangian. We stress that this observation only holds when the theory is written in terms of  $\psi_a, \chi^i$ and $\rho=2\lambda+\gamma^a\psi_a$, i.e. the dilatino $\lambda$ only occurs  through the generalized dilatino $\rho$.

This conclusion can also be readily reached from ${\cal N}=1$ SDFT, where it follows from the fact that the vector index of the gravitino transforms under  O(1,9)$_R$  and the spinor index under  O(9,1)$_L$, reflecting its  R-NS (or NS-R) origin in the string spectrum.

Moreover, since the transformations \eqref{transfe} preserve the number of dilatini, $\beta$ invariance  can be separately analyzed on  three generic combinations of terms, namely
\ba
{\cal L}_0&=&F_5+a_1F_6+a_2B_3^F+a_3B_2^H+a_4B_4^H+a_5B_6^H\, ,\\
{\cal L} _1&=&F_8+b_1F_{10}+b_2B_3^\phi+b_3B_1^F+b_4B_8^H \, ,\\
{\cal L}_2&=&F_1+cB_1^H\, .
\ea 
These terms transform as
\ba
\delta_\beta{\cal L}_0&=& \ (1 - 2 a_1)  \beta_{ac}A^{ci}\left(  \ov\psi^{a} \gamma^{b} \nabla_{b} \chi_{i}-\nabla_{b}\ov\psi^{a} \gamma^{b}\chi_{i}\right)
\cr
&&- \ 
  (1+ 2 a_5 )\left[2 \beta_{cd} (w^{d}{}_{ab}+\frac12H^d{}_{ab})
-   \beta_{bd} w^{d}{}_{ac}+\frac12\beta_{bd}\left(H^d{}_{ac}+A^d{}_{i}F_{ac}^i\right)\right] \ \ov \psi^{a} \gamma^{b} \psi^{c}\cr 
&& 
-
\ (1+2 a_2 -2 a_5 )\beta_{c}{}^{d}  A_{di} F_{ab}{}^{i}  \ \ov \psi^{a} \gamma^{b} \psi^{c} +\   \vphantom{\frac12}(2 a_1 + a_2) \beta^{bc}F_{ab}{}^{i} A_{c}{}^{j} \ov\chi_{i} \gamma^{a} \chi_{j} \cr
&&+\frac18(a_5+24a_3)\left[\left(H_{dab}+A_{di}F_{ab}^i\right)\beta_c{}^d-\nabla_a\beta_{bc}\right]\ov\chi_j\gamma^{abc}\chi^j\cr
&&+\ \frac18
(1 + 24 a_4)\left[\vphantom{\frac12}\right.
\left( H_{ecd} + A_{ei} F_{cd}{}^{i}\right) \beta_{b}{}^{e} 
-  \nabla_{d}\beta_{bc} 
\left.\vphantom{\frac12}\right] \ov \psi^{a}\gamma^{bcd} \psi_{a}\cr
&&- \left[\vphantom{\frac12}\right. 
 (2 a_5 - a_2)  H_{dab}  A_{c}{}^{i} \beta^{cd}
+ (1+ 2 a_1 + 2 a_2)  \left( A_{c}{}^i w_{dab} \beta^{cd} 
- F_{bc}{}^{i} \beta_{a}{}^{c} \right) 
\left.\vphantom{\frac12}\right]\ov\psi^{a}\gamma^{b} \chi_{i} \cr
 &&+ \ 2 (a_4 - 2 a_3 )  A^{e}{}_{i}  H_{bcd} \beta_{ae}
 \ov\psi^{a}\gamma^{bcd} \chi^{i}\, ,
\end{eqnarray}
\ba
\delta_\beta{\cal L}_1&=& \ (1+\frac12 b_2 )   \beta^{cd}  H_{acd} \ \ov\rho \psi^{a} +\frac18(1-b_1+16b_3-8b_4)A_{di}F_{bc}^i\beta_a{}^d \ov\psi^{a} \gamma^{bc} \rho\cr
&&+\ 
 \frac18  (1-b_1+8b_4)
\left[\vphantom{\frac12}(2
  w_{d}{}_{bc}+
 H_{dbc} )\beta_{a}{}^{d}
+2 \left(2w_{dab}-
 H_{dab} - A_{di} F_{ab}{}^{i}\right)\beta_{c}{}^{d}
\right] \ov\psi^{a} \gamma^{bc} \rho \cr
&&-\ \frac14 
\left[\vphantom{\frac12}\right.
(1-b_1+8b_3) w_{acd} - 4(b_3 - b_4)  H_{acd} 
\left.\vphantom{\frac12}\right]\beta^{ab}  A_{b}{}^{i}\ov\chi_{i} \gamma^{cd}\rho \, ,   \ea
\ba
\delta_\beta {\cal L}_2 &=& 
-\ \frac18 \left(1+24c \right)  \; 
\left[\nabla_{a}\beta_{bc}  - \left( H_{dab}+ A_{d}{}^{i} F_{ab i}\right) \beta_{c}{}^{d} \right]\overline{\rho}\gamma^{abc} \rho\, .
\ea

Hence,
 $\beta$ symmetry fixes all but four  coefficients, and the most general $\beta$ invariant combination is
\begin{eqnarray}
&& \alpha_1 \left( F_1 - \frac{1}{24} B^{H}_{1} \right) \;+\; \alpha_2 \left(F_{8} + F_{10} -2 B_{3}^{\phi}\right) \;+\; \alpha_{3} \left(8 \; F_{10} + B_1^{F} + B_8^H\right) \cr
&&
\;\;\;\;\;\;\;\;\;\;\;\;\;\;\;\;\;\;\;\;\;\;\;\;\;\;
+\; \alpha_4\left(F_5 + \frac12 F_6 - B_3^F -\frac{1}{48} B_{2}^{H}-\frac{1}{24} B_4^H -\frac12 B_{6}^H\right)\, ,
\end{eqnarray}
leading to 
\ba
S_{\rm F}&=&\int d^{10}xee^{-2\phi}\left[\alpha_1\left( \ov{\rho}\gamma^{a}\nabla_{a}\rho-\frac1{24}\ov{\rho}\gamma^{abc}\rho H_{abc}\right)+\ \alpha_2\left(\ov{\rho}\nabla^{a}\psi_{a}+ \ov{\psi}_{a}\nabla^{a}\rho-2\ov{\rho}\psi^{a}\nabla_{a}\phi\right)\right.\nn\\
&&\ \ \ \ \ \ \ \ \ \ \ \ \ \ \ \ \  \ +\ \alpha_{3}\left(8\ov{\psi}_{a}\nabla^{a}\rho+ \ov{\rho}\gamma^{ab}\chi_{i}F_{ab}{}^{i}+ \ov{\rho}\gamma^{ab}\psi^c H_{abc}\right)\nn\\
&&\ \ \ \ \ \ \ \ \ \ \ \ \ \ \ \ \ \ +\ \alpha_4\left( \ov{\psi}_a\gamma^{b}\nabla_{b}\psi_a+\frac1{2} \ov{\chi}_i\gamma^{a}\nabla_{a}\chi^i- \ov{\psi}^a\gamma^{b}\chi_{i}F_{ab}{}^{i}-\frac1{48}\ov{\chi}^i\gamma^{abc}\chi_i H_{abc}\right.\nn\\
&&\ \ \ \ \ \ \ \ \ \ \ \ \ \ \ \ \ \ \ \ \ \ \ \ \ \ \left.-\frac1{24} \ov{\psi}_d\gamma^{abc}\psi^d H_{abc}-\frac12 \ov{\psi}^a\gamma^{b}\psi^{c}H_{abc}\right)\, .
\ea

The undetermined  coefficients
can be fixed by supersymmetry. Parameterizing \eqref{gentransf0}, we get the following supersymmetry transformation rules 
\be
\delta_\epsilon e_{\mu}{}^{a} = \frac{1}{2}\ov{\epsilon}\gamma^{a}\psi_{\mu}\, , \qquad \delta_\epsilon e^{\mu}{}_{a} = - \frac{1}{2}\ov{\epsilon}\gamma^{\mu}\psi_{a}\, , \qquad \delta_\epsilon\phi = - \frac{1}{4}\ov{\epsilon}\rho + \frac{1}{4}\ov{\epsilon}\gamma^{a}\psi_{a}\, ,
\ee
\be
\delta_\epsilon b_{\mu\nu} = \ov{\epsilon}\gamma_{[\mu}\psi_{\nu]} + \frac{1}{2}\ov{\epsilon}\gamma_{[\mu}\chi^{m}A_{\nu]m}\, , \qquad \delta_\epsilon A_{\mu}{}^{m} = \frac{1}{2}\ov{\epsilon}\gamma_{\mu}\chi^{m}\, ,
\ee
\be
\delta_\epsilon\rho = \gamma^{a}\nabla_{a}\epsilon - \frac{1}{24}H_{abc}\gamma^{abc}\epsilon - \nabla_{a}\phi\gamma^{a}\epsilon\, , \qquad \delta_\epsilon\psi_{a} = \nabla_{a}\epsilon - \frac{1}{8}H_{abc}\gamma^{bc}\epsilon\, ,
\ee
\be
\delta_\epsilon\chi^{m} = - \frac{1}{4}F_{ab}{}^{m}\gamma^{ab}\epsilon\, ,
\ee
and  supersymmetry fixes
\be
\alpha_1=-\alpha_4=4\alpha_{3}\, .
\ee
Note that the terms with coefficient $\alpha_2$ add up to a total derivative. Hence, we are left with
\ba
S_{\rm F}&=&\alpha_1\int d^{10}xee^{-2\phi}\left[\ov{\rho}\gamma^{a}\nabla_{a}\rho-\frac1{24}\ov{\rho}\gamma^{abc}\rho H_{abc}\right.+\ \frac14\left(8\ov{\psi}_{a}\nabla^{a}\rho+ \ov{\rho}\gamma^{ab}\chi_{i}F_{ab}{}^{i}+ \ov{\rho}\gamma^{ab}\psi^c H_{abc}\right)\nn\\
&&\ \ \ \ \ \ \ \ \ \ \ \ \ \ \ \ \ \ -\ \left( \ov{\psi}_a\gamma^{b}\nabla_{b}\psi_a+\frac1{2} \ov{\chi}_i\gamma^{a}\nabla_{a}\chi^i- \ov{\psi}^a\gamma^{b}\chi_{i}F_{ab}{}^{i}-\frac1{48}\ov{\chi}^i\gamma^{abc}\chi_i H_{abc}\right.\nn\\
&&\ \ \ \ \ \ \ \ \ \ \ \ \ \ \ \ \ \ \ \ \ \ \  \left. \left.-\frac1{24} \ov{\psi}_d\gamma^{abc}\psi^d H_{abc}-\frac12 \ov{\psi}^a\gamma^{b}\psi^{c}H_{abc}\right)\right]+\ \alpha_2\int d^{10}xe\nabla_{a}\left(\ov{\rho}\psi^{a}e^{-2\phi}\right)\, .\nn\\ \label{sf}
\ea
Finally, considering the bosonic action $S_{\rm B}$ in
\eqref{sb}, supersymmmetry also fixes $\alpha_1=1$ and the sum
\be
S=S_{\rm B}+S_{\rm F}\, ,\label{totac}
\ee
exactly matches the action  of ${\cal N}=1$ supergravity coupled to super Yang-Mills  \cite{Bergshoeff:1981um},  up to total derivatives, field redefinitions and quartic fermion fields. In particular, it agrees with the  heterotic supergravity  action presented in \cite{bdr} applying the field redefinitions $\phi^{-3}\rightarrow e^{-2\phi}$, $R\rightarrow -R$, $H_{\mu\nu\lambda}\rightarrow\frac{1}{3\sqrt{2}}H_{\mu\nu\lambda}$, $B_{\mu\nu}\rightarrow\frac{1}{\sqrt{2}}b_{\mu\nu}$,  $\lambda\rightarrow\frac{1}{2\sqrt{2}}(\rho-\gamma^\mu\psi_\mu)$, $A_\mu\rightarrow \frac{1}{\sqrt{2}}A_\mu$, $\chi\rightarrow \frac{1}{\sqrt{2}}\chi$.

\section{Quartic fermion interactions} \label{four}

In this section we construct all possible $\beta$ invariant combinations of quartic fermion terms. Note that at leading order in $\alpha'$, these terms do not contain derivatives.

From \eqref{bitransf}, we see that $\beta$ invariance only allows contractions  of the vector index of a gravitino  with that of  another gravitino. These  contracted gravitini can be taken to be in the same bilinear using the Fierz rearrangement identity 
\be
\left(\ov\psi M\varphi\right)\left(\ov\eta N\chi\right)=-\frac1{32}\sum_{n=0}^5C_n\left(\ov\psi \gamma^{c_1...c_n}\chi\right)\left(\ov\eta N\gamma_{c_1...c_n}M\varphi\right)\, ,\label{fr}
\ee
where $C_0=C_1=2, C_2=-1, C_3=-\frac13, C_4=\frac1{12}, C_5=\frac1{120}$ and $\psi, \varphi, \eta, \chi$ are arbitrary Majorana-Weyl fermions. 
 Then, there is a scheme in which gravitini only appear in the structure $\ov\psi^d\gamma^{abc}\psi_d$, since bilinear couplings $\ov\eta\gamma^{c_1\cdots c_n}\eta$ of Majorana-Weyl spinors $\eta$ are  only non-vanishing for $n=3$. 

A similar analysis applies for gaugini. Since gauge invariance only allows contractions  between themselves\footnote{Contractions with $F_{ab}^i$  in quartic fermion terms are of higher order in $\alpha'$. }, using \eqref{fr} one can always find a scheme where they only appear in the structure $\ov\chi^i\gamma^{abc}\chi_i$.

In the scheme in which the vector indices of the gravitini  only contract among themselves and gaugini  only gauge contract with gaugini, dilatini can only come in pairs. However, \eqref{fr} implies that the unique non-vanishing quartic dilatino coupling $(\ov\rho\gamma^{abc}\rho)(\ov\rho\gamma_{abc}\rho)$ vanishes, and then only one factor $\ov\rho\gamma^{abc}\rho$ is allowed in a quartic fermion coupling.

In summary,  only three building blocks are allowed for quartic fermion couplings at leading order in $\alpha'$: $(\ov\rho\gamma^{abc}\rho), (\ov\chi^i\gamma^{abc}\chi_i)$ and $(\ov\psi^d\gamma^{abc}\psi_d)$. 

The $\beta$ transformations of these building blocks
\ba
\delta_\beta\left(\ov\rho\gamma^{abc}\rho\right)&=&3\beta^{d[a}\ov\rho\gamma^{bc]}{}_{d}\rho\, ,\label{tr}\\
\delta_\beta\left(\ov\chi^i\gamma^{abc}\chi_i\right)&=&-4\beta^{de}A_e{}^i\ov\chi_i\gamma^{abc}\psi_d+3\beta^{d[a}\ov\chi_i\gamma^{bc]}{}_{d}\chi_i\, ,\\
\delta_\beta\left(\ov\psi^d\gamma^{abc}\psi_d\right)&=&2\beta^{de}A_e{}^i\ov\chi_i\gamma^{abc}\psi_d+3\beta^{d[a}\ov\psi_e\gamma^{bc]}{}_d\psi^e\, ,\label{tp}
\ea
suggest the following definitions
\be
Y^{abc}\equiv\ov\chi^i\gamma^{abc}\chi_i+2\ov\psi^d\gamma^{abc}\psi_d\qquad{\rm and}\qquad Z^{abc}\equiv\ov\rho\gamma^{abc}\rho\, ,
\ee
which allow to condense \eqref{tr}-\eqref{tp} as
\be
\delta_\beta Y^{abc}=3\beta^{d[a}Y^{bc]}{}_d\qquad{\rm and}\qquad \delta_\beta Z^{abc}=3\beta^{d[a}Z^{bc]}{}_d\, .
\ee

It is straightforward to verify that  $Y^2$ and $Y\cdot Z$ are the only non-vanishing $\beta$ invariant quartic fermion couplings, in terms of the previous three building blocks.

We stress that this conclusion is implied by $\beta$ symmetry of supergravity. Nevertheless, it  can also be readily obtained from  ${\cal N}=1$ SDFT, where
one can only construct two quartic couplings of generalized fermion fields, namely
\be
\left(\ov\rho\gamma_{\un{abc}}\rho\right)\left(\ov\Psi^{\ov D}\gamma^{\un{abc}}\Psi_{\ov D}\right)\qquad{\rm and}\qquad \left(\ov\Psi^{\ov D}\gamma^{\un{abc}}\Psi_{\ov D}\right)\left(\ov\Psi^{\ov E}\gamma_{\un{abc}}\Psi_{\ov E}\right)\, . \label{2c}
\ee
 This is easy to see   using the SDFT version of \eqref{fr}, i.e.
\be
\left(\ov\Psi M\Phi\right)\left(\ov\Gamma N\Theta\right)=-\frac1{32}\sum_{n=0}^5(-1)^nC_n\left(\ov\Psi \gamma^{\un c_1...\un c_n}\Theta\right)\left(\ov\Gamma N\gamma_{\un c_1...\un c_n}M\Phi\right)\, ,\label{frdft}
\ee
where the factor $(-1)^n$ is due to the Clifford algebra \eqref{clif}.
Using the parameterizations \eqref{psi} and \eqref{rho}, in terms of supergravity and super Yang-Mills fields, the couplings \eqref{2c} are precisely $Y\cdot Z$ and $Y^2$.

To fix the relative coefficient between these two terms in the action, we appeal to supersymmetry.  
 Then, we must determine the cubic fermion corrections to the supersymmetry transformations in \eqref{gentransf0}. To that aim, we consider the fermion dependent piece of the action of ${\cal N}=1$ SDFT 
\be
S_f=\int d^{20+n_g}x \  e^{-2d} \left(L_{2f}+L_{4f}\right)\, ,
\ee
with
\ba
L_{2f}&=& \overline{\Psi}^{\overline{A}}\gamma^{\underline{b}}\nabla_{\underline b}{\Psi}_{\overline{A}}-\bar{\rho}\gamma^{\underline{a}}\nabla_{\underline{a}}\rho+2\overline{\Psi}^{\overline{A}}\nabla_{\overline{A}}\rho\, , \label{lf}\\
L_{4f}&=& \alpha_{1}\left(\ov{\rho}\gamma^{\un{abc}}\rho\right)\left(\ov{\Psi}{}^{\ov{D}}\gamma_{\un{abc}}\Psi_{\ov{D}}\right) + \alpha_{2}\left(\ov{\Psi}{}^{\ov{D}}\gamma^{\un{abc}}\Psi_{\ov{D}}\right)\left(\ov{\Psi}{}^{\ov{E}}\gamma_{\un{abc}}\Psi_{\ov{E}}\right)\, ,
\ea
where $L_{2f}$ was introduced in \cite{hk2}, and  the most general supersymmetry transformations involving cubic fermion fields 
\ba
\delta^{(3)}_{\epsilon}\Psi_{\ov{A}} & = & \beta_{1}\ov{\epsilon}\rho\Psi_{\ov{A}}  + \beta_{2}\ov{\epsilon}\gamma^{\un{bcde}}\rho\gamma_{\un{bcde}}\Psi_{\ov{A}}+ \beta_{3}\ov{\epsilon}\gamma^{\un{b}}\Psi_{\ov{A}}\gamma_{\un{b}}\rho + \beta_{4}\ov{\Psi}_{\ov{A}}\rho\epsilon\, ,\label{gen_gravitino}\\
\delta_{\epsilon}^{(3)}\rho&=& \zeta_{1}\ov{\rho}\gamma^{\un{abc}}\rho\gamma_{\un{abc}}\epsilon + \zeta_{2}\ov{\Psi}{}^{\ov{A}}\gamma^{\un{bcd}}\Psi_{\ov{A}}\gamma_{\un{bcd}}\epsilon\, . \label{gen_dilatino}
\ea
Note that   terms  with more than one bilinear fermion field in the supersymmetry transformations of the boson fields are higher order in $\alpha'$. 

Performing the variation of $S_f$,  supersymmetry fixes all the coefficients, so that
\be
L_{4f}= -\frac1{384}\left(\ov{\rho}\gamma^{\un{abc}}\rho\right)\left(\ov{\Psi}{}^{\ov{D}}\gamma_{\un{abc}}\Psi_{\ov{D}}\right) +\frac1{192}\left(\ov{\Psi}{}^{\ov{D}}\gamma^{\un{abc}}\Psi_{\ov{D}}\right)\left(\ov{\Psi}{}^{\ov{E}}\gamma_{\un{abc}}\Psi_{\ov{E}}\right)\, ,
\ee
and 
\ba
\delta^{(3)}_{\epsilon}\Psi_{\ov{A}} & = & -\frac14\ov{\epsilon}\rho\Psi_{\ov{A}}  +\frac14\ov{\epsilon}\gamma^{\un{b}}\Psi_{\ov{A}}\gamma_{\un{b}}\rho + \frac14\ov{\Psi}_{\ov{A}}\rho\epsilon\, ,\label{gen_gravitino}\\
\delta_{\epsilon}^{(3)}\rho&=& \frac1{384}\ov{\rho}\gamma^{\un{abc}}\rho\gamma_{\un{abc}}\epsilon + \frac1{96}\ov{\Psi}{}^{\ov{A}}\gamma^{\un{bcd}}\Psi_{\ov{A}}\gamma_{\un{bcd}}\epsilon\, . \label{gen_dilatino}
\ea

Parameterizing \eqref{gen_gravitino} and \eqref{gen_dilatino} with \eqref{param}-\eqref{parrho} leads to the following  cubic fermion contributions to the supersymmetry transformation rules of the standard supergravity fields 
\ba
\label{gen_gravitino2}
\delta^{(3)}_{\epsilon}\psi_{{a}} & = & -\frac14\ov{\epsilon}\rho\psi_{{a}} -\frac1{4}\ov{\epsilon}\gamma^{{b}}\psi_{a}\gamma_{{b}}\rho +\frac14\ov{\psi}_{a}\rho\epsilon\, ,\\
\delta^{(3)}_{\epsilon}\chi_{i} & = &- \frac14\ov{\epsilon}\rho\chi_{i} -\frac1{4}\ov{\epsilon}\gamma^{{b}}\chi_{i}\gamma_{{b}}\rho +\frac14\ov{\chi}_{{i}}\rho\epsilon\, ,\\
\delta_{\epsilon}^{(3)}\rho&=&- \frac1{384}\ov{\rho}\gamma^{{abc}}\rho\gamma_{{abc}}\epsilon -\frac1{96}\ov{\psi}{}^{{a}}\gamma^{{bcd}}\psi_{a}\gamma_{{bcd}}\epsilon-\frac1{192}\ov{\chi}{}^{i}\gamma^{{bcd}}\chi_{i}\gamma_{{bcd}}\epsilon\, ,
\ea
and  quartic fermion Lagrangian 
\be
-\frac1{48\times 16}{\sqrt{-g}\ e^{-2\phi}}\left(Y_{abc}Y^{abc}-Y_{abc}Z^{abc}\right)\, .\label{qf}
\ee
 We show in  Appendix \ref{apb}  that this simple expression exactly agrees with the quartic fermion interactions obtained in \cite{Bergshoeff:1981um,bdr}, where they are spread through various covariant derivatives and gauge curvatures. The same simple form of the Lagrangian and supersymmetry transformation rules were obtained in \cite{strick} using generalized geometry.

The simplicity of this expression highlights the power of $\beta$ symmetry as an organizing principle and suggests that it can provide a useful tool to compute higher derivative couplings.

\section{Conclusions}\label{conclu}

In this paper we explored  $\beta$ symmetry in heterotic supergravity, extending the results of \cite{bmn,bmn2,by} to the massless gauge and fermionic sectors.

The $\beta$ transformation rules of the  massless heterotic fields  were obtained from  ${\cal N}=1$ SDFT, by gauge fixing  the double  O(9,1) $\times$ O(1, 9+$n_g$)  Lorentz group to the single O(1,9) Lorentz group  of supergravity. Although the   framework of SDFT  is convenient, we stress that $\beta$ symmetry acts at the level of the standard supergravity fields, with no need to introduce dual coordinates or to appeal to alternative formulations of supergravity. Actually,  the non-geometric sector of O($d,d$) parameterized by the bi-vector $\beta$ acts as a hidden symmetry in standard ten dimensional supergravity, unlike the so-called $\beta$-supergravity  scheme    \cite{andriot,andriot2}, in which the non-geometric sector is realized by the shifts of the Kalb-Ramond field, or the generalized geometric scheme of \cite{coimbra,strick}, which requires  the mathematical structure of Courant algebroid.

We would like to emphasize that ten dimensional supergravity is not invariant under O($10,10$), and then $\beta$ transformations are not  symmetries  of the ten dimensional theory. However, its compactification on $T^n$ is invariant under O($n,n$), and  hence $\beta$ invariant. The condition $\beta^{\mu\nu}\partial_\nu...=0$ constrains the field configurations to not depend on the
coordinates along which $\beta$ has non-zero components. Therefore, it  implies that one is effectively dealing with the compactified action, in a scheme that preserves the ten dimensional structures.   

While $\beta$ symmetry completely fixes the interactions of the boson fields,  we found  four $\beta$ invariant combinations of  bilinear fermion
couplings and two of quartic fermion fields, whose relative coefficients in the action must be fixed by supersymmetry. The source of the less predictive power of $\beta$ symmetry in the fermionic sector as compared to that of the bosonic one, lies in the fact that the transformations preserve
 the number of generalized dilatini  in contrast to those for dilatons (compare equations \eqref{transferho} and \eqref{nrho}  with \eqref{transfedil}, \eqref{fi2} and \eqref{sfi}). Nevertheless, $\beta$ invariance turned out to be a fruitful organizing principle, not only  by allowing an independent treatment of terms containing different  numbers of generalized dilatini, but also by excluding contractions of the form $\gamma^a\psi_a$ in the scheme in which the fermion fields are $\psi_a, \rho$ and $\chi^i$ .

Beyond the obvious importance of identifying new symmetries, determining $\beta$ symmetry in the gauge and fermionic sectors of heterotic supergravity might be relevant to establish the duality structure of the low energy limit of string theory when higher derivative terms are included. Indeed, although several methods have been used to compute higher orders  and important progress has been made in recent years,  the duality structure of the $\alpha'$ expansion is unknown yet.  In particular,   the tree level $\zeta(3)\alpha'^3$ terms that are common to all string theories  were
completely determined  using T-duality symmetry in \cite{garou}, but they
have a very complicated structure. Even if some terms could be simplified  imposing O($d,d$) symmetry in the dimensional reduction of the pure gravitational interactions \cite{wulff}, there seem to be obstructions to a manifest duality covariant formulation of the full expression  \cite{hw}.  We postpone a detailed analysis of  $\beta$ symmetry in presence of higher derivative corrections to  a forthcoming paper, but a  few comments to appreciate its relevance in that context might be useful here.

Symmetries  play an important role in the calculation of  higher-derivative corrections in  the string effective field theories. But they hold iteratively in powers of $\alpha'$,
so that the transformation rules of the  fields demand order by order modifications, which must  be  additionally consistent with other  symmetries and dualities of string theory.  Explicitly, the interplay between supersymmetry and O($d,d$) symmetry   has been the subject of several  papers \cite{waldram} $-$ \cite{liu}.  In particular, the form of
the first order deformations of the supersymmetry transformation rules of the  heterotic  fields obtained from a manifestly duality covariant formalism  in \cite{lnr}\footnote{See also \cite{liu} where the toroidal reduction of the supersymmetry variations  of the heterotic fermion fields at ${\cal O}(\alpha')$ agree with the expressions in \cite{lnr} when the double Lorentz group is gauge fixed to the single Lorentz group of supergravity.} differs from that obtained by supersymmetrizing the Chern-Simons forms  in \cite{bdr}. Furthermore, it has been argued that there must be several supersymmetric (Riemann)$^4$ invariants \cite{lium,wulff}, but not all of them will be compatible with string dualities.  However, since the string effective field theories are defined up to field redefinitions, which  may even contain unconventional  non-covariant gauge terms \cite{mn}, the  comparison among different schemes of the theory is not obvious. To verify the compatibility between supersymmetry and  O($d,d$) symmetry, the ten dimensional theory must be compactified.  $\beta$ symmetry emerges as a useful tool in this framework, since $\beta$ transformations act covariantly on the multiplets of the ten dimensional diffeomorphisms. Hence, it allows to avoid   the field redefinitions that are required  by the usual Kaluza Klein procedure.

Likewise, while $\beta$ transformations are unique to lowest order,   they are also deformed by higher derivative corrections. 
To first order in $\alpha'$,
$\beta$ invariance of the  universal NS-NS sector  was proved in \cite{bmn,bmn2}.
The deformations of the $\beta$ transformation rules, closure of the symmetry algebra and  invariance of the Lagrangian were presented  in the so-called generalized Bergshoeff-de Roo,  Metsaev-Tseytlin  and DFT schemes, which are related by field redefinitions.
Regarding the fermionic sector,   such as at leading order, we do not expect  $\beta$ symmetry to completely fix the action at higher orders.  Furthermore, although $\beta$  invariance restricts the fermion couplings  at leading order, the  restrictions  can be circumvented  at higher orders by field redefinitions. Actually, even if the  contraction of the  vector index of  the gravitini with that of the $\gamma$ matrices  is not allowed in  the DFT scheme of supergravity  at any order,  in other schemes it might be introduced by field redefinitions. Specifically,  terms proportional to the field equations of motion were added  in equation (3.17) of \cite{bdr},  which contain a contraction $\gamma^a\psi_a$ that cannot be absorbed in the generalized dilatino. Nevertheless, at the next order in $\alpha'$,  the most general form of the action required for supersymmetrization of the Chern-Simons forms, namely equation (4.6) in $loc.cit.$, can be rewritten in terms  of $\rho$ without further  $\gamma^a\psi_a$ factors. Moreover, the absence of terms involving the structure $\gamma^a\psi_a$ is also verified in the manifestly duality invariant form of the ${\cal O}(\alpha')$ corrections presented in \cite{lnr}.

Another direction for which the formulation  developed in this paper may prove useful  is the study of  integrable  deformations of two dimensional $\sigma$ models.
The last years have seen significant progress in 
 understanding   the connection between the  bi-vector $\beta$  and the classical Yang-Baxter equation as a solution generating technique in (generalized) supergravity  (see \cite{Sakamoto:2017cpu}$-$\cite{Osten:2023cza} and references therein).  In particular, our approach might be suitable for  generalizations of the  deformations explored in the gauge sector of the heterotic string  \cite{Osten:2023cza},   including  supersymmetry.

\bigskip

{\bf Acknowledgements:} We warmly thank D. Marqu\'es for useful comments and discussions. Support by Consejo Nacional de Investigaciones Cient\'ificas y T\'ecnicas (CONICET), Agencia Nacional de Promoci\'on Científica y T\'ecnica (ANPCyT), Universidad de La Plata (UNLP) and Universidad de Buenos Aires (UBA) is gratefully acknowledged. 

\bigskip

\begin{appendix}
\section{Notation and definitions}\label{AppA}
We use  $\mu,\nu,\rho,\dots$ and  $a,b,c,\dots$ indices for space-time and tangent space coordinates, respectively. The infinitesimal Lorentz transformation of the vielbein is
\be
\delta_\Lambda e_\mu{}^a = e_\mu{}^b \Lambda_b{}^a \ \, .
\ee
The  spin connection 
\be
w_{c a b} = -e^\mu{}_c\left(\partial_\mu e_{\nu a}e^\nu{}_b -\Gamma^\rho_{\mu\nu} e_{\rho a} e^\nu{}_b\right)\, ,\qquad {\rm with }\ \ \ \ \  \Gamma^\rho_{\mu\nu}=\frac12 g^{\rho\sigma}\left(\partial_\mu g_{\sigma\nu}+\partial_\nu g_{\mu\sigma}-\partial_\sigma g_{\mu\nu}\right)\, ,
\ee
transforms as
\be
\delta_\Lambda w_{c a b} =  D_c \Lambda_{a b} + w_{d a b}  \Lambda^d{}_c  + 2 w_{c d [b} \Lambda^d{}_{a]}\, ,
\ee
and hence, it turns flat derivatives $D_a$ into covariant flat derivatives $\nabla_a$ as
\be
\nabla_a T_b{}^c = D_a T_b{}^c+ w_{a b}{}^d T_d{}^c-w_{ad}{}^cT_b{}^d \ , \ \ \ D_a = e^\mu{}_a \partial_\mu\, .
\ee 
The Christoffel connection $\Gamma^\rho_{\mu\nu}$ turns spacetime  partial into covariant derivatives as
\be
\nabla_\mu T_{\rho}{}^\sigma
=\partial_\mu T_{\rho}{}^\sigma-\Gamma_{\mu\rho}^\lambda T_\lambda{}^\sigma +\Gamma_{\mu\lambda}^\sigma T_\rho{}^\lambda\, .
\ee

The Riemann tensor 
\be
R^\mu{}_{\nu\rho\sigma}=\partial_\rho\Gamma^\mu_{\nu\sigma}-\partial_\sigma\Gamma^\mu_{\nu\rho}+\Gamma^\mu_{\rho\lambda}\Gamma^\lambda_{\nu\sigma}-\Gamma^\mu_{\sigma\lambda}\Gamma^\lambda_{\nu\rho}\, ,
\ee
with flat spacetime indices is defined as
\be
R_{a b c d} = 2 D_{[a}w_{ b] c d} + 2 w_{[a b]}{}^e w_{e c d} + 2 w_{[{a} |c|}{}^e w_{{b}] e d}\, .
\ee
While the symmetry $R_{a b c d} = R_{[ab] [cd]}$ is manifest, other symmetries of the Riemann tensor are hidden and  determine the Bianchi identities
\be
R_{a b c d} = R_{c d a b} \ , \ \ \  \ R_{[a b c] d} = 0 \\ , \ \ \  \ \nabla_{[a}R_{b c] d e} = 0  . \label{BI2}
\ee
The Ricci tensor and scalar curvature are given by the traces
\be
R_{a b} = R^c{}_{a c b} \ , \ \ \  R = R_a{}^a \ .  \ \ \ 
R_{[a b]} = 0\ . \label{BI3}
\ee

\subsection{Useful $\gamma$-identities}
We have  used the following identities of $\gamma$ matrices:
\ba
\left[\gamma^{ab},\gamma_{c_1\cdots c_n}\right]&=&
4n\delta^{[b}_{[c_1}\gamma^{a]}{}_{c_2\cdots c_n]}\label{i1} \;,\\
\gamma^b\gamma_{cde}\gamma^a&=&-\gamma^a\gamma_{cde}\gamma^b-2g^{ab}\gamma_{cde}-24\delta^a_{[c}\gamma_d\delta^b_{e]}+6\delta^a_{[c}\gamma_{de]}\gamma^b+6\gamma^a\gamma_{[cd}\delta^b_{e]}\;,\label{i2}\\
\gamma_{abc} \gamma^{d} + 2\gamma^{d} \gamma_{abc} &=&
24 \gamma_{[ab}\delta_{c]}^{d} - 2\gamma^{d} \gamma_{abc}\, ,\label{i3} \\
4 \gamma_{abc} \gamma^{d} + \gamma^{d} \gamma_{abc} &=&
-3 \gamma^{d}{}_{abc} + 15 \delta^{d}_{[a} \gamma_{bc]}\label{i4} \;.
\ea

\section{Comparing quartic fermion interactions }\label{apb}

In this Appendix we show that the quartic fermion interactions \eqref{qf} determined from $\beta$ symmetry agree with those obtained  in \cite{Bergshoeff:1981um} from supersymmetry.  We use the notation employed in \cite{bdr},  except for the replacement $\Gamma\rightarrow\gamma$.

The quartic fermion terms in  equation (B.6) of reference \cite{bdr}, are
\begin{eqnarray}
{\cal{L}}_{BdR}(R)&=& e^{-2d} \frac{1}{48} \psi^{d}\gamma^{abc}\psi_{d}\left(
\bar{\lambda}\gamma_{abc} \lambda 
+ \frac{\sqrt{2}}{2} \bar\lambda \gamma_{abc} \gamma^{d}\psi_{d} 
- \frac14 \bar\psi^{d}\gamma_{abc} \psi_{d} -\frac18 \bar\psi^{d} \gamma_{d}\gamma_{abc} \gamma^{e} \psi_{e} \right) . \;\;\;\; \;\;\;\; \label{LR}
\end{eqnarray}
In terms of the generalized dilaton $\rho=2\sqrt2\lambda+\gamma^d\psi_d$, this becomes
\begin{eqnarray}
{\cal{L}}_{BdR}(R)&=&  \frac{e^{-2d}}{48\times 8} \psi^{d}\gamma^{abc}\psi_{d}\left(
\bar{\rho}\gamma_{abc} \rho
- 2 \bar\psi^{d}\gamma_{abc} \psi_{d} \right)  \;.
\end{eqnarray}
The quartic fermion terms in equation  (B.7) of \cite{bdr} are (we use $\beta=\frac12$)
\begin{eqnarray}
{\cal{L}}_{BdR}(F^2)&=& e^{-2d} \left[\vphantom{\frac12}\right.\frac{1}{8}  \ov\chi^{i} \gamma^{d}\gamma^{ab}\left(\psi_{d} + \frac{\sqrt{2}}{3} \gamma_{d} \lambda \right)\left( \ov\psi_{a}\gamma_{b}\chi_{i} \right)  - \frac{1}{16} \left(\ov\chi^{i}\gamma^{abc} \chi_{i} \right) \left(\psi_{a}\gamma_{b}\psi_{c}\right) \cr 
&&-\frac{\sqrt{2}}{16\times 24} \left(\ov\chi^{i}\gamma^{abc} \chi_{i} \right)\ov\psi_d
\left(4\gamma_{abc}\gamma^{d} + 3 \gamma^{d}\gamma_{abc}\right)\lambda
+\frac{1}{2\times 48} \left(\ov\chi^{i}\gamma^{abc} \chi_{i} \right)
\left(\bar\lambda \gamma_{abc} \lambda \right)\cr
&& -\frac{1}{16\times 48} \left(\ov\chi^{i}\gamma^{abc} \chi_{i} \right)
\left(\ov\chi^{j}\gamma_{abc} \chi_{j} \right) 
\left.\vphantom{\frac12}\right] \, ,\label{LF2}
\end{eqnarray}
where the first term is the fermionic contribution from $-\frac{1}{8}  \hat F^i_{ab} \ov\chi_i \gamma^{d}\gamma^{ab}\left(\psi_{d} + \frac{\sqrt{2}}{3} \gamma_{d} \lambda \right)$ with ${\hat F}^i_{ab} = F^i_{ab} -\ov\psi_{[a}\gamma_{b]} \chi^i$, and the second term is the contribution from $\frac{\sqrt2}{16}\ov\chi^i\gamma^{abc}\chi_i\hat H_{abc}-\frac12\ov \chi^i \slashed{D}(\omega(e,\psi)) \chi_i$ with ${\hat H}_{abc}=  H_{abc} - \frac{\sqrt{2}}{4} \ov\psi_{[a} \gamma_{b} \psi_{c]}$ and $\slashed{D}$  the standard covariant derivative with the replacement 
\begin{eqnarray}
\omega_{\mu ab}(e)\to\omega_{\mu ab}(e,\psi)=\omega_{\mu ab}(e)+\frac14 \left(\ov\psi_{\mu} \gamma_{a} \psi_{b} - \ov\psi_{\mu} \gamma_{b} \psi_{a} + \ov\psi_{a} \gamma_{\mu} \psi_{b}\right)\;.    
\end{eqnarray}
The first term in (\ref{LF2}) becomes
\begin{eqnarray}
\frac{1}{8} e^{-2d} \ov\chi^{i} \gamma^{d}\gamma^{ab}\left(\psi_{d} + \frac{\sqrt{2}}{3} \gamma_{d} \lambda \right)\left( \ov\psi_{a}\gamma_{b}\chi_{i} \right)=e^{-2d} \left[\vphantom{\frac12}\right.\frac{1}{8}  \left(\ov\chi^{i} \gamma^{ab} \rho\right) 
+ \frac{1}{2}  \left(\ov\chi^{i} \gamma^{[b}\psi^{a]} \right)\left.\vphantom{\frac12}\right]
\left( \ov\psi_{a}\gamma_{b}\chi_{i} \right) \, ,
\label{aux1}
\end{eqnarray}
and using the Fierz rearrangement formula (\ref{fr}), these two terms can be rewritten as
\begin{eqnarray}
\frac{1}{8}  \left(\ov\chi^{i} \gamma^{ab} \rho\right) \left(\ov\psi_{a}\gamma_{b} \chi_{i}\right) 
&=&
\frac{1}{48\times 16}  \left(\ov\chi^{i} \gamma^{abc} \chi_{i}\right) \left[-3 \left(\ov \psi^{d}\gamma_{dabc}\rho\right)
+ 15 \left(\ov \psi_{a}\gamma_{bc}\rho\right) \right]\, ,\label{aux1A}\\
 \frac{1}{2}  \left(\chi^{i} \gamma^{[b}\psi^{a]} \right)
\left( \bar\psi_{a}\gamma_{b}\chi_{i} \right) &=&
\left(\chi^{i} \gamma^{cde} \chi_{i}\right) \left[\vphantom{\frac12}\right.
\frac{1}{48\times 8} \bar\psi_{a} \gamma^{a} \gamma_{cde} \gamma^{b}\psi_{b}
- \frac{1}{48\times 4} \bar\psi^{a} \gamma_{cde} \psi_{a} \cr 
&& \;\;\;\;\;\;\;\;\;\;\;\;\;\;\;\;\;\;\;\;\;\;\;\;\;\;\;\;
+  \frac{1}{16} \bar\psi_{c} \gamma_{d} \psi_{e} 
- \frac{1}{32} \bar\psi_{c} \gamma_{de} \gamma^{a}\psi_{a}
 \left.\vphantom{\frac12}\right]\;,\label{aux1B}
\end{eqnarray}
where we used $C_3=-\frac13$, identity (\ref{i2}) and the fact that $\ov \chi^{i} \gamma_{c_1 \dots c_n}\chi_{i}$ vanishes for $n\neq3$.

In terms of the generalized dilatino, the second line of (\ref{LF2}) becomes
\begin{eqnarray}
\frac{1}{16\times 48}\left(\bar\chi^{i}\gamma^{abc} \chi_{i} \right)
\left[ 24 \bar\psi_{a}\gamma_{bc} \gamma^{e}\psi_{e} 
-2 
\bar\psi_{d}\gamma^{d} \gamma_{abc} \gamma^{e}\psi_{e} 
+3
\left( \bar\psi_{d}\gamma^{d}{}_{abc} -5 \bar\psi_{[a} \gamma_{bc]}\right)\rho 
+
\bar\rho \gamma_{abc} \rho  \right]\, ,
\label{aux2}
\end{eqnarray}
where we used the identities (\ref{i3}) and (\ref{i4}) .

Putting all together, we get 
\begin{eqnarray}
{\cal{L}}_{BdR}(F^2)&=& \frac{e^{-2d}}{16\times 48} \left(\bar\chi^{i}\gamma^{abc} \chi_{i} \right)\left(\vphantom{\frac12}\right.
\bar\rho\gamma_{abc} \rho 
- \bar\chi^{j}\gamma_{abc} \chi_{j}
 - 4 \bar\psi^{d}\gamma_{abc} \psi_{d} 
\left.\vphantom{\frac12}\right) \, ,\label{LF2Final}
\end{eqnarray}
 so that the quartic fermion couplings can be compactly written as
\begin{eqnarray}
L_{BdR}(R)+ L_{BdR}(F^2) &=& -\frac{e^{-2d}}{48\times 16} Y^{abc}\left(Y_{abc}- Z_{abc}\right)    \equiv -\frac{e^{-2d}}{3} X^{abc} X_{abc}  \, ,  \label{LFerm4}
\end{eqnarray}
where we defined
\begin{eqnarray}
Y^{abc}=\ov\chi^i\gamma^{abc}\chi_i+2\ov\psi^d\gamma^{abc}\psi_d\, ,\qquad Z^{abc}=\ov\rho\gamma^{abc}\rho\, ,\qquad
 X^{abc}=\frac{1}{16}\left(Y_{abc}-\frac12 Z_{abc}\right)  \;,
\end{eqnarray}
and  used the fact that $Z_{abc} Z^{abc}=0$.

\end{appendix}


\begin{thebibliography}{99} 
\bibitem{sen} A.~Sen,
``O(d) x O(d) symmetry of the space of cosmological solutions in string theory, scale factor duality and two-dimensional black holes,''
Phys. Lett. B \textbf{271} (1991), 295-300


\bibitem{bmn} W.~H.~Baron, D.~Marques and C.~A.~Nunez,
``\ensuremath{\beta} Symmetry of Supergravity,''
Phys. Rev. Lett. \textbf{130} (2023) no.6, 061601
[arXiv:2209.02079 [hep-th]].

\bibitem{bmn2}
W.~H.~Baron, D.~Marques and C.~A.~Nunez,
``Exploring the \ensuremath{\beta} symmetry of supergravity,''
JHEP \textbf{12} (2023), 006
[arXiv:2307.02537 [hep-th]].

\bibitem{by}
W.~H.~Baron and N.~A.~Yazbek,
``\ensuremath{\beta} symmetry in type II supergravities,''
JHEP \textbf{03} (2024), 146
[arXiv:2312.15061 [hep-th]].


\bibitem{hk2}  O. Hohm and S. K. Kwak, N=1 Supersymmetric Double Field Theory, JHEP
 \textbf{03} (2012), 080 [arXiv:1111.7293 [hep-th]].

\bibitem{Bergshoeff:1981um}
E.~Bergshoeff, M.~de Roo, B.~de Wit and P.~van Nieuwenhuizen,
``Ten-Dimensional Maxwell-Einstein Supergravity, Its Currents, and the Issue of Its Auxiliary Fields,''
Nucl. Phys. B \textbf{195} (1982), 97-136

\bibitem{jlp} 
I.~Jeon, K.~Lee and J.~H.~Park,
``Supersymmetric Double Field Theory: Stringy Reformulation of Supergravity,''
Phys. Rev. D \textbf{85} (2012), 081501
[erratum: Phys. Rev. D \textbf{86} (2012), 089903]
[arXiv:1112.0069 [hep-th]].


\bibitem{hk} O.~Hohm and S.~K.~Kwak,
``Double Field Theory Formulation of Heterotic Strings,''
JHEP \textbf{06} (2011), 096
[arXiv:1103.2136 [hep-th]].

\bibitem{cmnp} M. Ciafardini, D. Marqu\'es, C. N\'u\~nez and A. Pereyra Grau,
``Hidden symmetries from extra dimensions," arXiv:2410.07325

\bibitem{bdr} E.~A.~Bergshoeff and M.~de Roo,
``The Quartic Effective Action of the Heterotic String and Supersymmetry,''
Nucl. Phys. B \textbf{328} (1989), 439-468

\bibitem{strick}
J.~Kupka, C.~Strickland-Constable and F.~Valach,
``Direct derivation of gauged $\mathcal N=1$ supergravity in ten dimensions to all orders in fermions,''
[arXiv:2410.16046 [hep-th]].



\bibitem{andriot}
D.~Andriot and A.~Betz,
``$\beta$-supergravity: a ten-dimensional theory with non-geometric fluxes, and its geometric framework,''
JHEP \textbf{12} (2013), 083
[arXiv:1306.4381 [hep-th]].

\bibitem{andriot2} D.~Andriot and A.~Betz,
``NS-branes, source corrected Bianchi identities, and more on backgrounds with non-geometric fluxes,''
JHEP \textbf{07} (2014), 059
[arXiv:1402.5972 [hep-th]].

\bibitem{coimbra}
A.~Coimbra, C.~Strickland-Constable and D.~Waldram,
``Supergravity as Generalised Geometry I: Type II Theories,''
JHEP \textbf{11} (2011), 091
[arXiv:1107.1733 [hep-th]].


\bibitem{garou}
M.~R.~Garousi,
``Effective action of type II superstring theories at order $\alpha'^{3}$: NS-NS couplings,''
JHEP \textbf{02} (2021), 157
[arXiv:2011.02753 [hep-th]].



\bibitem{wulff} L.~Wulff,
``Completing R$^{4}$ using O(d,d),''
JHEP \textbf{08} (2022), 187
[arXiv:2111.00018 [hep-th]].

\bibitem{hw}
S.~Hronek and L.~Wulff,
``$O(D,D)$ and the string $\alpha'$ expansion: an obstruction,''
JHEP \textbf{04} (2021), 013
[arXiv:2012.13410 [hep-th]].


\bibitem{waldram}
A.~Coimbra, R.~Minasian, H.~Triendl and D.~Waldram,
``Generalised geometry for string corrections,''
JHEP \textbf{11} (2014), 160
[arXiv:1407.7542 [hep-th]].

\bibitem{Baron:2018lve}
W.~H.~Baron, E.~Lescano and D.~Marqu\'es,
``The generalized Bergshoeff-de Roo identification,''
JHEP \textbf{11} (2018), 160
[arXiv:1810.01427 [hep-th]].

\bibitem{lnr} E.~Lescano, C.~A.~N\'u\~nez and J.~A.~Rodr\'\i{}guez,
``Supersymmetry, T-duality and heterotic \ensuremath{\alpha}'-corrections,''
JHEP \textbf{07} (2021), 092
[arXiv:2104.09545 [hep-th]].

\bibitem{Butter:2021dtu}
D.~Butter,
``Exploring the geometry of supersymmetric double field theory,''
JHEP \textbf{01} (2022), 152
[arXiv:2101.10328 [hep-th]].

\bibitem{Butter:2022gbc}
D.~Butter,
``Type II double field theory in superspace,''
JHEP \textbf{02} (2023), 187
[arXiv:2209.07296 [hep-th]].



\bibitem{sezgin}
H.~Y.~Chang, E.~Sezgin and Y.~Tanii,
``Higher derivative couplings of hypermultiplets,''
JHEP \textbf{06} (2023), 172
[arXiv:2304.06073 [hep-th]].

\bibitem{Butter:2023nxm}
D.~Butter, F.~Hassler, C.~N.~Pope and H.~Zhang,
``Generalized dualities and supergroups,''
JHEP \textbf{12} (2023), 052
[arXiv:2307.05665 [hep-th]].

\bibitem{Liu:2023fqq}
J.~T.~Liu and R.~J.~Saskowski,
``Consistent truncations in higher derivative supergravity,''
JHEP \textbf{09} (2023), 136
[arXiv:2307.12420 [hep-th]].

\bibitem{lium}
J.~T.~Liu and R.~Minasian,
``Higher-derivative couplings in string theory: five-point contact terms,''
Nucl. Phys. B \textbf{967} (2021), 115386
[arXiv:1912.10974 [hep-th]].


\bibitem{liu}
S.~Jayaprakash and J.~T.~Liu,
``Higher derivative heterotic supergravity on a torus and supersymmetry,''
[arXiv:2406.14600 [hep-th]].


\bibitem{mn} D.~Marques and C.~A.~Nunez,
``T-duality and \ensuremath{\alpha}'-corrections,''
JHEP \textbf{10} (2015), 084
[arXiv:1507.00652 [hep-th]].




\bibitem{Sakamoto:2017cpu}
J.~i.~Sakamoto, Y.~Sakatani and K.~Yoshida,
``Homogeneous Yang-Baxter deformations as generalized diffeomorphisms,''
J. Phys. A \textbf{50} (2017) no.41, 415401
[arXiv:1705.07116 [hep-th]].


\bibitem{Fernandez-Melgarejo:2017oyu}
J.~J.~Fernandez-Melgarejo, J.~i.~Sakamoto, Y.~Sakatani and K.~Yoshida,
``$T$-folds from Yang-Baxter deformations,''
JHEP \textbf{12} (2017), 108
[arXiv:1710.06849 [hep-th]].

\bibitem{yb}
I.~Bakhmatov and E.~T.~Musaev,
``Classical Yang-Baxter equation from $\beta$-supergravity,''
JHEP \textbf{01} (2019), 140
[arXiv:1811.09056 [hep-th]].

\bibitem{yb2} R.~Borsato, A.~Vilar L\'opez and L.~Wulff,
``The first $\alpha'$-correction to homogeneous Yang-Baxter deformations using $O(d, d)$,''
JHEP \textbf{07} (2020) no.07, 103
[arXiv:2003.05867 [hep-th]].

\bibitem{yb3} D.~Butter, F.~Hassler, C.~N.~Pope and H.~Zhang,
``Consistent truncations and dualities,''
JHEP \textbf{04} (2023), 007
[arXiv:2211.13241 [hep-th]].

\bibitem{Osten:2023cza}
D.~Osten,
``Heterotic integrable deformation of the principal chiral model,''
Phys. Rev. D \textbf{109} (2024) no.10, 106021
[arXiv:2312.10149 [hep-th]].
















\end{thebibliography}
\end{document}